\documentclass[12pt]{article}
\usepackage{graphicx}
\input epsf
\def\hybrid{\topmargin 0pt      \oddsidemargin 0pt
        \headheight 0pt \headsep 0pt
        \voffset=-0.5cm
        \textwidth 6.25in       
        \textheight 9.5in       
        \marginparwidth 0.0in
        \parskip 5pt plus 1pt   \jot = 1.5ex}
\catcode`\@=11
\def\marginnote#1{}

\newcount\hour
\newcount\minute
\newtoks\amorpm
\hour=\time\divide\hour by60
\minute=\time{\multiply\hour by60 \global\advance\minute by-\hour}
\edef\standardtime{{\ifnum\hour<12 \global\amorpm={am}%
        \else\global\amorpm={pm}\advance\hour by-12 \fi
        \ifnum\hour=0 \hour=12 \fi
        \number\hour:\ifnum\minute<10 0\fi\number\minute\the\amorpm}}
\edef\militarytime{\number\hour:\ifnum\minute<10 0\fi\number\minute}

\def\draftlabel#1{{\@bsphack\if@filesw {\let\thepage\relax
   \xdef\@gtempa{\write\@auxout{\string
      \newlabel{#1}{{\@currentlabel}{\thepage}}}}}\@gtempa
   \if@nobreak \ifvmode\nobreak\fi\fi\fi\@esphack}
        \gdef\@eqnlabel{#1}}
\def\@eqnlabel{}
\def\@vacuum{}
\def\draftmarginnote#1{\marginpar{\raggedright\scriptsize\tt#1}}
\def\draftlabel#1{{\@bsphack\if@filesw {\let\thepage\relax
   \xdef\@gtempa{\write\@auxout{\string
      \newlabel{#1}{{\@currentlabel}{\thepage}}}}}\@gtempa
   \if@nobreak \ifvmode\nobreak\fi\fi\fi\@esphack}
        \gdef\@eqnlabel{#1}}
\def\@eqnlabel{}
\def\@vacuum{}
\def\draftmarginnote#1{\marginpar{\raggedright\scriptsize\tt#1}}

\def\draft{\oddsidemargin -.5truein
        \def\@oddfoot{\sl preliminary draft \hfil
        \rm\thepage\hfil\sl\today\quad\militarytime}
        \let\@evenfoot\@oddfoot \overfullrule 3pt
        \let\label=\draftlabel
        \let\marginnote=\draftmarginnote
   \def\@eqnnum{(\theequation)\rlap{\kern\marginparsep\tt\@eqnlabel}%
\global\let\@eqnlabel\@vacuum}  }

\def\numberbysection{\@addtoreset{equation}{section}
        \def\theequation{\thesection.\arabic{equation}}}

\def\underline#1{\relax\ifmmode\@@underline#1\else
        $\@@underline{\hbox{#1}}$\relax\fi}

\def\titlepage{\@restonecolfalse\if@twocolumn\@restonecoltrue\onecolumn
     \else \newpage \fi \thispagestyle{empty}\c@page\z@
        \def\thefootnote{\fnsymbol{footnote}} }

\def\endtitlepage{\if@restonecol\twocolumn \else  \fi
        \def\thefootnote{\arabic{footnote}}
        \setcounter{footnote}{0}}  
\relax

\numberbysection
\hybrid

\newfont{\Bbb}{msbm10 scaled 1\@ptsize00}
\newfont{\Bbbb}{msbm7 scaled 1\@ptsize00}
\newcommand{\CC}{\mbox{\Bbb C}}

\newcommand{\DDD}{\raise-1pt\hbox{$\mbox{\Bbbb D}$}}

\newcommand{\UUU}{\raise-1pt\hbox{$\mbox{\Bbbb U}$}}

\newcommand{\z}{\raise-1pt\hbox{$\mbox{\Bbbb Z}$}}

\def\beq{\begin{equation}}
\def\eeq{\end{equation}}
\def\p{\partial}

\newtheorem{theorem}{Theorem}[section]
\newtheorem{lemma}{Lemma}[section]
\newtheorem{lemma-definition}{Lemma-Definition}[section]
\newtheorem{corollary}{Corollary}[section]

\newtheorem{proposition}{Proposition}[section]

\def\square{\hfill
{\vrule height6pt width6pt depth1pt} \break \vspace{.01cm}}

\begin{document}
\begin{titlepage}

\title{Symmetric solutions of the dispersionless Toda hierarchy
and associated conformal dynamics}

\author{A.~Zabrodin
\thanks{Institute of Biochemical Physics,
4 Kosygina st., Moscow 119334, Russia; ITEP, 25
B.Cheremushkinskaya, Moscow 117218, Russia and
National Research University Higher School of Economics,
20 Myasnitskaya Ulitsa,
Moscow 101000, Russia; e-mail: zabrodin@itep.ru}}

\date{June 2013}
\maketitle

\vspace{-7cm} \centerline{ \hfill ITEP-TH-19/13} \vspace{7cm}

\begin{abstract}

Under certain reality conditions, a general solution
to the dispersionless Toda lattice hierarchy describes
deformations of simply-connected plane domains with a smooth
boundary. The solution depends on an arbitrary (real positive)
function 
of two variables which plays the role of a density 
or a conformal metric in the plane.
We consider in detail the important class of 
symmetric solutions characterized by 
the density functions that depend only on the distance
from the origin and that are positive and regular in an annulus
$r_0< |z|<r_1$. We construct the dispersionless tau-function 
which gives formal local solution to the 
inverse potential problem and to the Riemann mapping problem
and discuss the associated conformal dynamics 
related to viscous flows in the Hele-Shaw cell.

\end{abstract}

\vfill

\end{titlepage}
\newpage

\tableofcontents

\newpage

\section{Introduction}

In papers \cite{MWWZ}-\cite{KMZ05}
it was shown that some classical problems 
of complex analysis in 2D, such as 
the inverse potential problem, the Dirichlet boundary value problem 
and related problems of viscous hydrodynamics
have a hidden integrable
structure. For simply-connected domains with a smooth 
enough boundary, it
is the 2D Toda lattice (2DTL) hierarchy of Ueno-Takasaki 
\cite{UenoTakasaki}
in the limit of zero dispersion \cite{TakTak} while
in more general cases it is the universal Whitham hierarchy
introduced by Krichever
in \cite{KriW1,KriW,KriW2}. 

In the hydrodynamical context, this integrable structure 
applies to viscous flows in the 
Hele-Shaw cell with negligible surface tension and, more generally, 
to a class of growth problems referred to 
as Laplacian growth (LG). They appear in different physical and
mathematical contexts and have many important applications
(see, e.g., \cite{RMP}-\cite{MWPT} and references therein). 
In the 2D LG processes, the dynamics of a moving front or interface
between two distinct phases (a closed curve in the plane) is driven by
a harmonic scalar field in the domain bounded by the curve.

The hierarchical times $t_k, \bar t_k$ ($k\geq 1$) of the 2DTL 
hierarchy are suitably
normalized harmonic moments of the domain and their complex
conjugates, with $t_0$ being proportional
to the area of the domain. The
dispersionless tau-function $F=F(t_0,\{t_k\},
\{\bar t_k\})$, regarded as a function 
of the moments, is a kind of the master
function for the above mentioned problems.
In particular, it contains all the information
about the conformal bijection of any domain with given moments
to the unit disk. 

The function $F$ is a particular solution to 
the dispersionless version
of the Hirota equations for the 2DTL hierarchy.
Although it admits a simple and explicit integral representation,
it is a highly non-trivial function when regarded as a function 
of the $t_k$'s.
Some recurrence combinatorial formulas for coefficients of its 
Taylor expansion are available \cite{N03,KKN}.

Further, in \cite{Ztmf,Z07} it was argued that any 
{\it non-degenerate}
solution of the hierarchy, with certain
reality conditions imposed, can be given a
similar geometrical and hydrodynamical
meaning\footnote{{\it Degenerate} solutions
(known also as finite-component reductions of the 
infinite hierarchy) were shown in \cite{GT} to be related to
conformal maps of slit domains.}.
Such solutions are parameterized by
a function $\sigma (z, \bar z)$
of two variables which has the meaning of a 
background charge density, or conformal metric,
in the complex plane. The moments should be now defined as integrals
of powers of $z$ with this density. The integral representation for the
dispersionless tau-function also changes accordingly but the formulas
which express the conformal map through its second order derivatives
do not depend on the choice of 
$\sigma$. In other words, the Toda dynamics encodes
the shape dependence of the conformal mapping, which we call the
{\it conformal dynamics}.

In \cite{NZ13}, an important class of solutions to the 
dispersionless 2DTL hierarchy was distinguished.
These solutions are characterized by the property that the derivatives
$\p F/\p t_k$ restricted to the line $t_1=t_2=t_3=\ldots =0$ 
and $t_0\neq 0$ are zero for all $k\geq 1$.
In the context of the
conformal dynamics, they correspond to axially
symmetric functions $\sigma$ (i.e., the ones depending only on $|z|^2$).
The corresponding dispersionless tau-functions 
admit recurrence combinatorial formulas 
for coefficients of their 
Taylor expansion which generalize those obtained in \cite{N03}.

An important example 
($\sigma = R/|z|^2$) was considered in \cite{Z12}. It was shown
that, on one hand, this solution describes the LG on the surface 
of an infinite cylinder of radius $R$ (or in a channel with periodic
boundary conditions) and, on the other hand, it is closely related to 
enumerative 
algebraic geometry of ramified coverings of Riemann surfaces.
Namely, the dispersionless tau-function $F$ for this solution
is a generating function for the double Hurwitz numbers which count
connected genus 0 coverings of the Riemann sphere with prescribed
ramification type at two points. See \cite{Lando} for a review
of the subject and \cite{Okounkov00}-\cite{AMMN11} for various 
integrable properties 
of Hurwitz partition functions.

This paper is devoted to a more detailed exposition of 
symmetric solutions and their meaning in conformal dynamics. 
Contrary to the previous works, where the function 
$\sigma$ was almost always assumed 
to be regular in the whole plane except maybe 
at infinity, our assumptions here are much weaker.
We systematically study the 
case when the function $\sigma$ is only assumed to be regular 
in an annulus ${\sf A}=\{ z\in \CC \, |\, r_0 <|z|<r_1\}$ with
some $r_0, r_1$, and the boundary curve is within this annulus.
In fact this does not bring any substantial changes
because really important is only the local behavior of the
function $\sigma$ 
in a small neighborhood of the boundary curve.
However, some formulas get modified in this more general setting
because not all quantities remain to be well-defined and thus
require a more accurate definition.
As a result, some 
quantities may acquire dependence on 
the auxiliary parameter $r_0$ which, in physical terms, plays 
the role of a short-distance cut-off
for divergent integrals.
For example, the dispersionless tau-function 
for the domain ${\sf D}$ should be defined as
\beq\label{Faa}
F=-\, \frac{1}{\pi^2}
\int \!\! \int_{{\sf A}\cap {\sf D}} \int \!\! \int_{{\sf A}\cap {\sf D}}
\sigma (z, \bar z)\log \left |z^{-1} - \zeta^{-1}\right |
\sigma (\zeta , \bar \zeta )\, d^2 z d^2\zeta .
\eeq

We also give a number of explicit examples
containing all previously studied cases as well as some new ones.
In fact all these examples belong to a rather general family
\beq\label{family}
\displaystyle{\sigma (z, \bar z) \propto 
\frac{1}{z\bar z}\,(C_1 \log (z\bar z)+C_0)^{-\frac{k-3}{k-2}}}
\eeq
with integer $k>2$ and some constants
$C_0, C_1$. At $k\to 2$ with properly adjusted $C_0, C_1$ one gets
the family of homogeneous densities $\sigma (z, \bar z) \propto 
(z\bar z)^{\alpha -1}$. Among them are the cases $\sigma =1$
($\alpha =1$) discussed in \cite{MWWZ,WZ,KKMWZ} and 
$\sigma \propto 1/|z|^2$ ($\alpha =0$) discussed recently in
\cite{NZ13,Z12}. The latter is also the $k=3$ case of the 
general family (\ref{family}).

The paper is organized as follows. In section 2 we 
review the theory in the general (not necessarily symmetric) case,
with the modifications caused by the cut-off at $|z|=r_0$.
In section 3 we give a detailed analysis in the case of symmetric background 
densities $\sigma$. We 
generalize some familar results to non-zero values of $r_0$ and also
present some
statements and formulas which seem to be
absent in the literature (Theorem \ref{lambda} and
Corollaries \ref{lambdaj}, \ref{cor:hom}).
The explicit examples
are given in section 3.3.

\section{Deformations of plane domains and dispersionless
integrability}

\vspace{1ex}

The generic solutions to the
dispersionless 2DTL hierarchy take on a geometric significance when
the Toda times $t_k$, $\bar t_k$ are identified with (complex conjugate)
moments of simply-connected
domains in the complex plane with smooth boundary.
In this case the Toda
dynamics encodes the shape dependence
of the conformal mapping of such a domain to some
fixed reference domain.

\subsection{Local coordinates in the space of simply-connected domains}

Let ${\sf D}\subset \CC$ be a compact simply-connected
domain whose boundary is a smooth curve $\gamma=\partial {\sf D}$ and let
${\sf D^c}=\hat{\CC} \setminus {\sf D}$ be its complement in the
Riemann sphere $\hat{\CC}$.

\begin{figure}[tb]
\epsfysize=5.0cm
\centerline{\epsfbox{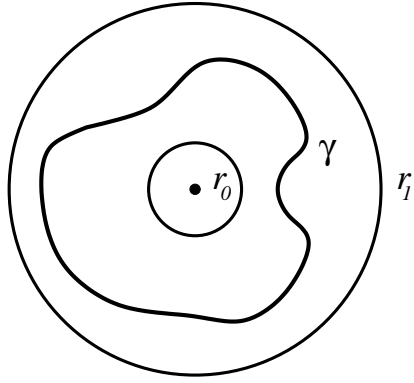}}
\caption{\sl The annulus ${\sf A}=
\{z\in \CC |\, r_0\leq |z|\leq r_1\}={\sf B}(r_1)\setminus {\sf B}(r_0)$
and the curve $\gamma \subset {\sf A}$.}
\label{fi:annulus}
\end{figure}

Let ${\sf B}(r)=\Bigl \{z\in \CC \Bigr |
\, |z|\leq r\Bigr \}$ be the disk of radius $r$ centered at the
origin. Without loss of generality, we assume that
${\sf B}(r_0)\subset {\sf D}$ and ${\sf D}\subset {\sf B}(r_1)$
for some $r_0<r_1$, i.e. the curve $\gamma$ belongs to
the annulus ${\sf A}={\sf B}(r_1)\setminus {\sf B}(r_0)$.
We will consider deformations of the domain ${\sf D}$
such that its boundary remains in the annulus
(Fig. \ref{fi:annulus}).

Fix a real-analytic and real-valued function $U(z, \bar z)$
in ${\sf A}$ such that
$$\sigma (z, \bar z):=\partial\bar\partial U(z, \bar z)>0,
\quad z\in {\sf A}$$
(we write $\p :=\p /\p z, \bar \p :=\p / \p \bar z$).
The function $\sigma$ plays the role of a
background charge density in the complex plane and
the function $U$ is the electrostatic potential created
by these charges.
We introduce the set of harmonic moments as follows:
\beq\label{loc1}
t_k=\frac{1}{2\pi ik}\oint_{\gamma} z^{-k}\partial U (z, \bar z) \, dz
\,,\quad k\geq 1.
\eeq
Using the Green's theorem, this contour integral can be
represented as a 2D integral over ${\sf B}(r_1)\setminus {\sf D}=
{\sf A}\cap {\sf D^c}$
(or ${\sf D} \setminus {\sf B}(r_0)={\sf A}\cap
{\sf D}$) plus a domain-independent
contour integral over $\p {\sf B}(r_1)$:
\beq\label{loc2}
t_k=-\, \frac{1}{\pi k}\int \!\!\!
\int_{{\sf A}\cap {\sf D^c}}\! z^{-k}
\sigma  (z, \bar z) \, d^2 z  \,  +\,
\frac{1}{2\pi ik}\oint_{|z|=r_1}z^{-k}\partial U (z, \bar z) \, dz,
\eeq
where $d^2z \equiv dx dy$.
In general $t_k$'s are complex numbers.
We claim that together with the real parameter
\beq\label{loc3}
t_0 = \frac{1}{2\pi i}\oint_{\gamma}\partial U (z, \bar z) \, dz
=\frac{1}{\pi }\int \!\!\! \int_{{\sf A}\cap {\sf D}}\!
\sigma  (z, \bar z) \,  d^2 z +
\frac{1}{2\pi i}\oint_{|z|=r_0}\partial U (z, \bar z) \, dz
\eeq
(the moment of constant function)
they form a set of local coordinates in the space
of domains ${\sf D}$.

This means, first, that any small deformation of a given
domain that preserves
all its moments is trivial
(local uniqueness of a domain with
given moments \cite{EV,KMZ05}).

\begin{proposition}\label{local-uniqueness}
Any one-parameter deformation ${\sf D}(t)$
of ${\sf D}={\sf D}(0)$ with some real parameter $t$
such that all $t_k$ are preserved, $\p_t t_k=0$, $k\geq 0$, is trivial.
\end{proposition}

\noindent
The proof is a modification
of the one presented in \cite{KMZ05} for the case $\sigma (z, \bar z)=1$.
We omit the proof because it is almost literally the same as
in \cite{NZ13}, where it was assumed, in addition,
that $\sigma$ is a regular
function in the whole plane. But this assumption is actually
irrelevant because what really matters is
the behavior of $\sigma$ in a vicinity
of the curve. In fact it is enough
to require that $\sigma$ is
regular and $\sigma \neq 0$ in some strip-like neighbourhood
of $\gamma$.

Second, the set of moments is not overcomplete, i.e., they
are independent parameters.
This fact follows
from the explicit construction of vector fields in the space of
domains that change real or imaginary part of any moment keeping
all the others fixed (see below). These arguments allow one to prove
the following theorem.

\begin{theorem}\label{local-coordinates}
The real parameters $t_0$, ${\rm Re}\, t_k$, ${\rm Im}\, t_k$,
$k\geq 1$, form a set of local coordinates in the space of simply-connected
plane domains with smooth boundary.
\end{theorem}

\noindent
This statement allows one to identify functionals on the space
of domains ${\sf D}$ with functions of infinitely many independent
variables $t_0 , \{ t_k \}, \{ \bar t_k \}$.

\subsection{The Green's function and
special deformations}

\subsubsection{The Green's function and the Poisson formula}

According to the Riemann mapping theorem, there exists
a conformal map $w(z)$ from ${\sf D^c}$ onto the
exterior of the unit circle. It is convenient to normalize it
by the conditions
$w(\infty )=\infty$ and $w'(\infty )$ is real positive.
The Laurent expansion of $w(z)$ at infinity has the form
$\displaystyle{
w(z)=pz+\sum\limits_{j\geq 0} p_j z^{-j}}
$,
where $p>0$.

If the conformal map $w(z)$ is known,
one can construct the Green's function of the
Dirichlet boundary value problem
in the domain ${\sf D^c}$:
\begin{equation} \label{Green}
G(z,\xi)=
\log\left\vert\frac{w(z)-w(\xi)}{w(z)
\overline{w(\xi)}-1}\right\vert .
\end{equation}
This function solves the Dirichlet boundary value problem
through the use of the Poisson formula
\beq\label{Poisson}
u^H(z)=-\frac{1}{2\pi}\oint_{\gamma}u(\xi)\partial_{n_{\xi}}
G(z,\xi)|d\xi|.
\eeq
Here $\partial_{n_{\xi}}$ denotes the derivative along the
outward normal vector to the boundary of ${\sf D}$
with respect to the second variable and
$|d\xi|$ is an infinitesimal element of length along the boundary.
The Poisson formula
provides the (unique) harmonic continuation
$u^H$ of any function $u$ from the curve $\gamma$ to its exterior
(i.e., the harmonic function in ${\sf D^c}$ regular at $\infty$
such that $u^H\bigl |_{\gamma}=u$).

The Green's function has the following properties: a)
it is symmetric under permutation of
the arguments, b) it is harmonic in each variable everywhere
in ${\sf D^c}$ except $z=\xi$ where it has the
logarithmic singularity $G(z,\xi)=
\log|z-\xi|+ \ldots $ as $z\to \xi$, c) $G(z,\xi)=0$
for any $z\in {\sf D^c}$ and $\xi \in \gamma$.

\subsubsection{Special deformations induced by the Green's function}

We will describe infinitezimal deformations of the domain
${\sf D}$ by the normal displacement of the
boundary $\delta n(z)$ at any point $z\in \gamma$,
positive if directed
outward ${\sf D}$.

Fix a point $a\in {\sf D^c}$ and consider a special
infinitesimal deformation
defined by the normal displacement
\beq\label{special}
\delta_a n (z)=-\frac{\varepsilon}{2\sigma (z, \bar z)}\,
\p_{n_z} G(a,z)\,, \quad \quad z\in \gamma, \,\,\, \varepsilon \to 0.
\eeq
Equivalently,
one may speak about the normal ``velocity''
of the boundary deformation
which is $V_n(z)=\mbox{lim}_{\varepsilon \to 0}
(\delta n_a(z)/\varepsilon )$, with $\varepsilon$ playing
the role of time.
Note that
$\p_{n_z} G(a,z)<0$, so at positive $\varepsilon$ the
domain expands.
For any sufficiently smooth initial
boundary this deformation is well-defined as $\varepsilon \to 0$.
By $\delta_a$ we denote the variation of any quantity under this deformation.

Let us introduce the differential operator
\beq\label{nabla}
\nabla(z) =\partial_0 + D(z)+ \bar D(\bar z),
\eeq
where $D(z), \bar D(\bar z)$ are given by
\beq\label{DD}
D(z)=\sum_{k\geq 1}\frac{z^{-k}}{k}\, \p_{k}\,,
\quad \quad
\bar D(\bar z)=\sum_{k\geq 1}\frac{\bar z^{-k}}{k}\, \bar \p_{k}\,.
\eeq
Hereafter we abbreviate $\p_k =\p /\p t_k$, $\bar \p_k =\p/\p \bar t_k$.

\begin{lemma}\label{change}  Let $X$ be any functional on the space
of domains ${\sf D}$ regarded as a function of $t_0, \{ t_k\},
\{\bar t_k\}$, then
for any $z\in {\sf D^c}$ we have $\delta_z X =\varepsilon \nabla(z)X$.
\end{lemma}

\noindent
{\it Proof.} From (\ref{loc2}), (\ref{loc3}) it is easy to see that
$$
\delta_z t_0 =-\frac{\varepsilon}{2\pi}
\oint_{\gamma} \p_{n_{\xi}} G(z, \xi ) |d\xi |=\varepsilon\,,\quad
\delta_z t_k =-\frac{\varepsilon}{2\pi k}
\oint_{\gamma} \xi^{-k}\p_{n_{\xi}} G(z, \xi ) |d\xi |=
\frac{\varepsilon}{k}\, z^{-k}
$$
by virtue of the Poisson formula (\ref{Poisson}).
Therefore by Theorem \ref{local-coordinates} we have:
$$
\delta_z X=\frac{\partial X}{\partial t_0}\, \delta_z t_0 +
\sum_{k\geq 1}\frac{\partial X}{\partial t_k}\, \delta_z t_k +
\sum_{k\geq 1}\frac{\partial X}{\partial \bar t_k}\, \delta_z \bar t_k=
\varepsilon
\Bigl (\partial_0 + \sum_{k\geq 1}\frac{z^{-k}}{k}\, \partial_k+
\sum_{k\geq 1} \frac{\bar{z}^{-k}}{k}\, \bar{\partial}_k\Bigr )X.
$$
\vspace{-5mm}
\square

\vspace{-3mm}

\begin{lemma}\label{harmonic}
Let $X$ be a functional of the form $X=
\int_{{\sf A}\cap {\sf D}}\Psi (\zeta , \bar \zeta )
\, \sigma (\zeta , \bar \zeta ) \, d^2\zeta$ with an
arbitrary domain-independent integrable
function $\Psi$ regular on the boundary, then
$$
\nabla (z)X= \pi \Psi ^H (z),
$$
where $\Psi^H(z)$ is the (unique) harmonic extension of the function
$\Psi$ from the boundary to the domain ${\sf D^c}$.
\end{lemma}

\noindent
{\it Proof.} The variation of $X$ under the special deformation
(\ref{special}) is
$$\delta_z X=\oint_{\gamma}\Psi (\zeta , \bar \zeta )
\sigma (\zeta , \bar \zeta )\delta n_z (\zeta )\, |d\zeta |=
-\frac{\varepsilon}{2}  \oint_{\gamma}\Psi (\zeta , \bar \zeta )
\p_{n_{\zeta}}G(z, \zeta )\, |d\zeta |.
$$
The assertion obviously follows from Lemma \ref{change} and
the Poisson formula (\ref{Poisson}).
\square

Now we can explicitly define the deformations that change only
either $x_k={\rm Re}\, t_k$ or $y_k={\rm Im}\, t_k$ keeping all
other moments fixed. From the proof of Lemma \ref{change} it follows that
the normal displacements $\delta n(\xi )=\varepsilon
{\rm Re}\, (\p_{n_{\xi}} H_k(\xi ))$ and
$\delta n(\xi )=\varepsilon
{\rm Im}\, (\p_{n_{\xi}} H_k(\xi ))$, where
$$
H_k(\xi )=\frac{1}{2\pi i}\oint_{\infty}z^k \p_z G(z, \xi )\, dz
$$
(the contour integral goes around infinity) change the real and
imaginary parts of $t_k$ by $\pm \varepsilon$ respectively
keeping all
other moments unchanged.
In particular, the deformation
\beq\label{deltainf}
\delta_{\infty}n(\xi )=-
\frac{\varepsilon }{2\sigma (\xi , \bar \xi )}\, \p_{n_{\xi}}
G(\infty , \xi )=\frac{\p_{n}\log |w(\xi )|}{2\sigma (\xi , \bar \xi )}
\eeq
changes $t_0$ only.
Therefore, the vector fields
$\p/ \p t_0$, $\p /\p x_k$, $\p / \p y_k$ in the space of
domains are locally well-defined and commute. Existence of such vector
fields means that the variables $t_k$ are independent and
$\p_k = \frac{1}{2}(\p_{x_k}-i\p_{y_k})$,
$\bar \p_k = \frac{1}{2}(\p_{x_k}+i\p_{y_k})$ can be understood
as partial derivatives.

\subsubsection{Deformations of the domain with given moments
induced by small changes of the potential}

Given a variation of the potential
$U\to U +\delta U$, one can consider a simultaneous
deformation of the boundary curve $\gamma$
such that all the moments $t_k$ remain fixed.

\begin{proposition}\label{deltaU} (cf. \cite{Z06})
Let $U\to U +\delta U$ be a variation of the potential, then
the deformation of the domain given by
\beq\label{delta1}
\delta n(z)=-\, \frac{\p_n
\bigl (\delta U(z, \bar z)\! -\! \delta U^H(z,
\bar z)\bigr )}{4 \sigma (z, \bar z)}
\eeq
preserves all the moments $t_k$, $k\geq 0$.
\end{proposition}

\noindent
{\it Proof.} The proof is straightforward.
At $k\geq 1$ we write (see (\ref{loc2})):
$\pi k \, \delta t_k = -\delta I_1 +\delta I_2$,
where
$$
\delta I_1 =\delta \left (\int \!\! \int_{{\sf A}\cap {\sf D^c}}
\!\! z^{-k}\sigma d^2 z\right )=-
\oint_{\gamma} z^{-k}\sigma \delta n(z)|dz|+
\int_{{\sf A}\cap {\sf D^c}} \!\! z^{-k}\delta \sigma d^2 z
$$
$$
= \frac{1}{4}\oint_{\gamma} \! z^{-k}
\p_n \!\! \left (\delta U(z, \bar z)\! -\! \delta U^H(z,
\bar z)\right )|dz|+\frac{1}{4}
\int \!\! \int_{{\sf A}\cap {\sf D^c}} \!\! z^{-k}\Delta \!
\left (\delta U(z, \bar z)\! -\! \delta U^H(z,
\bar z)\right )d^2z.
$$
Here $\Delta =4\p\bar \p$ is the Laplace operator and
$\delta U^H$ in the last integral can be added because
$\Delta (\delta U^H)=0$. By the Green theorem, the sum of the
two integrals in the last line yields
$$
\delta I_1 =\frac{1}{4}\oint_{|z|=r_1} \!\!
\left [z^{-k}\p_n
\! \left (\delta U \! -\! \delta U^H \right )
-\left (\delta U \! -\! \delta U^H \right )\p_n (z^{-k})\right ]|dz|
=\frac{1}{2i}\oint_{|z|=r_1} \!\! z^{-k}\p
\delta U \, dz
$$
(the last equality comes out as a result of some simple
transformations).
But this is equal to $\displaystyle{
\delta I_2 =\frac{1}{2i}\oint_{|z|=r_1} \!\! z^{-k}\p
\delta U \, dz}$.
Therefore, $\delta t_k =0$ for $k\geq 1$.
A similar calculation for $t_0$ (\ref{loc3}) gives
$\delta t_0=0$.
\square

\subsection{Complimentary moments}

The set of {\it complimentary moments} can be introduced
by the contour integrals
\beq\label{comp1}
v_k=\frac{1}{2\pi i}\oint_{\gamma} z^{k}\partial U (z, \bar z) \, dz \,,\quad k\geq 1.
\eeq
In the same way as in (\ref{loc2}), we can represent them in the form
\beq\label{comp2}
v_k=\frac{1}{\pi}\int \!\!\int_{{\sf A}\cap {\sf D}}\! z^{k}
\sigma  (z, \bar z) \, d^2 z   \, +\,
\frac{1}{2\pi i}\oint_{|z|=r_0} z^{k}\partial U (z, \bar z) \, dz.
\eeq
The moment $v_k$ is ``dual'' to the moment $t_k$ in the sense
which will be clarified below.
Dual to $t_0$ is the logarithmic moment
\beq\label{comp3}
v_0  =\frac{1}{\pi }\int \!\! \int_{{\sf A}\cap {\sf D}}\!
\log |z|^2 \sigma  (z, \bar z) \,  d^2 z
\eeq
which can be also represented through contour integrals:
\beq\label{comp3a}
v_0 =\frac{1}{2\pi i}\left (\oint_{\gamma}-\oint_{|z|=r_0}\right )
\Bigl ( \log |z|^2 \, \p U(z, \bar z) dz + U(z, \bar z)\,
d\log \bar z \Bigr ).
\eeq
The moments $v_k$ are functions of the moments
$t_0, \{t_k \}, \{\bar t_k\}$.

Consider the function
\beq\label{comp4}
\phi (z, \bar z)= -\, \frac{1}{\pi}\int \!\! \int_{{\sf A}\cap {\sf D}}
\log \left |z^{-1}-\zeta^{-1}\right |^2
\sigma (\zeta , \bar \zeta )\,
d^2\zeta
\eeq
which has the meaning of 2D Coulomb potential created by
the charge distributed in ${\sf A}\cap {\sf D}$ with
density $\sigma$ and a point-like charge at the origin.
This function is known to be continuous across the boundary
together with its first order partial derivatives:
if we write
$$
\phi (z, \bar z)=\Theta_{{\sf A}\cap {\sf D}}(z)
\phi^{(+)}(z, \bar z)\, + \,
\Theta_{{\sf D^c}}(z)
\phi^{(-)}(z, \bar z),
$$
where $\Theta_{{\sf D}}(z)$ is the characteristic
function of the domain
($\Theta_{{\sf D}}(z)=1$ if $z\in {\sf D}$ and
$0$ otherwise), then
\beq\label{cont}
\phi^{(+)}(z, \bar z)\Bigl |_{z\in \gamma} =
\phi^{(-)}(z, \bar z)\Bigr |_{z\in \gamma},
\quad \quad
\p \phi^{(+)}(z, \bar z)\Bigl |_{z\in \gamma} =
\p \phi^{(-)}(z, \bar z)\Bigr |_{z\in \gamma}.
\eeq
Equivalently, using the Green theorem, we can represent
the function $\phi$ as follows:
\beq\label{comp5}
\begin{array}{lll}
\phi (z, \bar z)&=&
-\, U(z, \bar z)\Theta_{{\sf A}\cap {\sf D}}(z)
\\ && \\
&&\displaystyle{
-\,\, \frac{1}{2\pi i}\left (\oint\limits_{\gamma}-
\oint\limits_{|z|=r_0}\right )
\left ( \log \bigl |z^{-1}-\zeta^{-1}\bigr |^2
\, \p_{\zeta} U(\zeta , \bar \zeta) d\zeta \, +\,
\frac{U(\zeta , \bar \zeta)\,\bar z}{(\bar \zeta -\bar z) \bar \zeta}\,
d\bar \zeta \right ).}
\end{array}
\eeq
In this form the continuity across the boundary is implicit
but the function $\phi$ becomes ready
for expanding it in a series both inside and outside
${\sf A}\cap {\sf D}$:
\beq\label{series1}
\phi^{(+)} (z, \bar z)=-\, U(z, \bar z)-u_0 +t_0\log |z|^2
+\sum_{k\geq 1}\bigl (t_k z^k + \bar t_k \bar z^k\bigr )+\psi (z, \bar z)\,,
\quad z\in {\sf A}\cap {\sf D}
\eeq
\beq\label{series2}
\phi^{(-)} (z, \bar z)=v_0 +\sum_{k\geq 1}
\frac{1}{k}\bigl ( v_k z^{-k}+\bar v_k \bar z^{-k}\bigr )+
\psi (z, \bar z)\,,
\quad \quad z\in {\sf D^c}
\eeq
Here
\beq\label{psi}
\psi (z, \bar z)=\frac{1}{2\pi i}\!
\oint_{|\zeta |=r_0}  \left [
\log \Bigl (1-\frac{\zeta}{z}\Bigr )\p_{\zeta}U d\zeta -
\log \Bigl (1-\frac{\bar \zeta}{\bar z}\Bigr )\p_{\bar \zeta}U d\bar \zeta
\right ]
\eeq
\beq\label{u0}
u_0=\frac{1}{2\pi i}\! \oint_{|\zeta |=r_0}
\left [\log |\zeta |^2 \p_{\zeta}U \, d\zeta +U d\log \bar \zeta
\right ].
\eeq
Note that $u_0$ is a real number and
$\psi$ is a harmonic function in $\CC \setminus
{\sf B}(r_0)$. This function
can be expanded in a series in negative powers of
$z$, $\bar z$ but here we do not need this expansion in the
explicit form.
We see that the moments $t_k$ determine the harmonic part
of the potential inside ${\sf A}\cap {\sf D}$
that depends on the shape of its exterior boundary $\gamma$
while the
complimentary moments are coefficients of the multipole expansion
of the potential outside it.

A holomorphic generating function for the moments is
given by the integral of Cauchy type
$$
C(z)=\frac{1}{2\pi i}\oint_{\gamma}
\frac{\p U(\zeta , \bar \zeta )\, d\zeta}{\zeta -z}\,.
$$
It defines a function holomorphic  in ${\sf D}$ and ${\sf D^c}$
with a jump across $\gamma$. Let $C^{\pm}(z)$ be the holomorphic
functions defined by this integral in ${\sf D}$ and ${\sf D^c}$
respectively. By the Sokhotski-Plemelj formula, the jump of the
function $C(z)$ across the contour $\gamma$
is equal to $\p U$:
\beq\label{jump}
(C^+(z)-C^-(z))\Bigr |_{z\in \gamma}= \p U(z, \bar z).
\eeq
This relation is nothing else than the equation of the
curve $\gamma$ in the $z$-plane.
Expanding $C^+(z)$ in the Taylor series for small
enough $|z|$, we see that it is the
generating function of the moments $t_k$ with $k\geq 1$:
$$
C^+(z)=\frac{1}{2\pi i}\oint_{\gamma}
\frac{\p U(\zeta , \bar \zeta )\, d\zeta}{\zeta -z}=
\sum_{k\geq 1}kt_kz^{k-1}\,.
$$
Similarly, $C^-(z)$ is the generating function for the
complimentary moments $v_k$ with $k\geq 1$:
$$
C^-(z)=\frac{1}{2\pi i}\oint_{\gamma}
\frac{\p U(\zeta , \bar \zeta )\, d\zeta}{\zeta -z}=-
\frac{t_0}{z}
-\sum_{k\geq 1}v_kz^{-k-1}, \quad z\in {\sf D^c}.
$$

\subsection{The dispersionless tau-function}

Consider the following functional on the space of domains ${\sf D}$:
\beq\label{F}
F=-\, \frac{1}{\pi^2}
\int \!\! \int_{{\sf A}\cap {\sf D}} \int \!\! \int_{{\sf A}\cap {\sf D}}
\sigma (z, \bar z)\log \left |z^{-1} - \zeta^{-1}\right |
\sigma (\zeta , \bar \zeta )\, d^2 z d^2\zeta .
\eeq
\begin{theorem} (cf. \cite{MWWZ}-\cite{KKMWZ},
\cite{Ztmf}) The following relations hold:
\beq\label{vk}
v_0=\p_0 F, \quad \quad v_k=\p_k F, \quad \quad \bar v_k=\bar \p_k F,
\quad \quad k\geq 1.
\eeq
\end{theorem}
\noindent
{\it Proof.}
The variation of $F$ under the special deformation
$\delta_z$ is
$$
\delta_z F= -\, \frac{\varepsilon}{4\pi} \oint_{\gamma}
\phi (\zeta , \bar \zeta )\p_{n_{\zeta}} G(z, \zeta )|d\zeta |
+\frac{1}{2\pi}\int \!\! \int_{{\sf A}\cap {\sf D}}
\delta_z \phi (\zeta , \bar \zeta )\sigma (\zeta , \bar \zeta )
d^2 \zeta \,.
$$
The first term is equal to $\varepsilon \phi (z , \bar z)/2$
by the Poisson formula with taking into account that
$\phi (z, \bar z)$ is harmonic in ${\sf D^c}$. In the second
term $\delta_z \phi (\zeta , \bar \zeta )=-\varepsilon
\log \left |z^{-1}-\zeta ^{-1}\right |^2$ because
the function $\log \left |z^{-1}-\zeta ^{-1}\right |^2$ for
$\zeta \notin {\sf D^c}$
is harmonic as a function of $z\in {\sf D^c}$ (see the proof of
Lemma \ref{harmonic}). Hence the second term is 
$\varepsilon \phi (z , \bar z)/2$, the same as the first one.
Therefore, by
Lemma \ref{change} we get
\beq\label{nablaF}
\nabla (z)F=
\phi (z, \bar z)=
-\frac{2}{\pi}\int \!\! \int_{{\sf A}\cap {\sf D}} \log
\left |z^{-1}-\zeta^{-1}\right |
\sigma (\zeta , \bar \zeta )\, d^2\zeta \,,
\quad z\in {\sf D^c}.
\eeq
The expansion of both sides in powers of $z$, $\bar z$ yields
(\ref{vk}).
\square

\vspace{-5mm}

\noindent
{\bf Remark.}
The assertion of the theorem means that the function $F$
gives a formal local solution to the inverse potential problem
in 2D: given $t_k$, one can find the complimentary moments 
by means of (\ref{vk}) and then the shape of the domain 
from the equation of the boundary curve (\ref{jump}).
The theorem also justifies the interpretation of the moment
$v_k$ as dual to $t_k$. In fact one could choose the $v_k$'s
as independent coordinates on the space of domains.
In these variables, one should work with the Legendre transformation
of the function $F$.

\vspace{-3mm}

\begin{theorem}\label{l4.2}(\cite{KKMWZ}-\cite{MWZ}, \cite{Ztmf}) It holds
\beq\label{GF}
G(z, \zeta )=\log \, \bigl | z^{-1} - \zeta ^{-1}\bigr |
+\frac{1}{2} \nabla (z) \nabla (\zeta )F.
\eeq
\end{theorem}

\noindent
{\it Proof.}
The proof consists in the application of Lemma \ref{harmonic}
to (\ref{nablaF})
and using the characteristic properties of the Green's function.
Applying the lemma to (\ref{nablaF}), we conclude that
$\nabla (\zeta )\nabla (z)F$ is the harmonic continuation of the
function $-2\log \left |z^{-1}-\zeta^{-1}\right |$ from the boundary
to the domain ${\sf D^c}$. This function is harmonic everywhere in
${\sf D^c}$ except at $\zeta =z$, where it has the logarithmic
singularity. It can be canceled, without changing
the boundary value, by adding the function $2G(z, \zeta )$.
\square

\vspace{-7mm}

\begin{corollary} \label{t4.1} The conformal map $w(z)$ is given by
\beq\label{w(z)}
w(z)=z\exp
\left (\Bigl (-\frac 12 \partial_0^2-\partial_0 D(z)\Bigr )F\right ).
\eeq
\end{corollary}

\noindent
{\it Proof.} From equation (\ref{Green}) it follows that
$G(z, \infty )=-\log |w(z)|$. Tending $\xi \to \infty$ in
(\ref{GF}) and separating holomorphic and antiholomorphic parts
in $z$, we get the result.
\square

\noindent
Note that the limit $z\rightarrow\infty$  in (\ref{w(z)})
yields
$\log p=-\frac{1}{2}\, \partial_0^2F.$

The following theorem establishes the connection
with integrability by identifying $F$ with the tau-function
of the dispersionless 2DTL hierarchy.

\begin{theorem} (\cite{MWWZ}-\cite{KKMWZ},
\cite{Ztmf})\label{t4.2}
The function $F$ satisfies the equations
\beq
\label{eqs}\begin{array}{crl}
(z-\zeta )e^{D(z)D(\zeta )F}&\!\! =\!\!&z e^{-\partial_0D(z)F}-\zeta e^{-\partial_0D(\zeta )F},
\\ &&\\
(\bar z-\bar\zeta )e^{\bar D(\bar z)\bar D(\bar \zeta )F}&\!\! =\!\! &
\bar z e^{-\partial_0\bar D(\bar z)F}-\bar\zeta
e^{-\partial_0\bar D(\bar\zeta )F},
\\ &&\\
1- e^{-D(z)\bar D(\bar\zeta )F}&\!\! =\!\! &
\displaystyle{(z\bar\zeta )^{-1} \,
e^{\partial_0 (\partial_0+D(z)+\bar D(\bar\zeta ))F}}.
\end{array}
\eeq
\end{theorem}

\noindent
The proof is the same as in \cite{KMZ05}.
Namely, combining (\ref{Green}) and (\ref{GF}), we get
$$\log\left\vert\frac{w(z)-w(\zeta )}{1-w(z)\bar w(\zeta )}
\right\vert= \log\left\vert\frac{1}{z} - \frac{1}{\zeta }\right\vert + \frac12\nabla(z)\nabla(\zeta )F.
$$
Next, substituting here $w(z)$ from (\ref{w(z)}) and separating
holomorphic and antiholomorphic parts
in $z$, $\zeta$, we obtain the result.

Equations (\ref{eqs}) comprise the dispersionless
2DTL hierarchy in the Hirota form \cite{TakTak}.
Note that although the definitions
of the harmonic moments and the function $F$
essentially depend on the background
density $\sigma$, the formulas for the Green's function and the
conformal map (\ref{GF}), (\ref{w(z)}) are $\sigma$-independent.
This means that the conformal dynamics is described by any solution
to the dispersionless 2DTL hierarchy of this class.

\begin{theorem}
The third order derivatives of the function $F$ are given by
\beq\label{3d}
\nabla (z_1)\nabla (z_2) \nabla (z_3) F=
-\, \frac{1}{4\pi}\oint_{\gamma}
\p_{n_{z}}\! G(z_1, z )\, \p_{n_{z}}\! G(z_2, z )\, 
\p_{n_{z}}\! G(z_3, z )\, \frac{|dz|}{\sigma (z ,\bar z)}\,.
\eeq
\end{theorem}

\noindent
{\it Proof.} This assertion  
follows from
the Hadamard variational formula \cite{Hadamard} for the 
Green's function:
$$
\delta G(z_1, z_2)=\frac{1}{2\pi}\oint_{\gamma}
\p_{n_{z}}\! G(z_1, z )\, \p_{n_{z}}\! G(z_2, z )\, \delta n(z) |dz|
$$
applied to the special deformation $\delta n (z)= \delta_{z_{3}}n(z)$
and from Lemma \ref{change}.
\square
\vspace{-5mm}

\noindent
For $\sigma (z, \bar z)=\mbox{const}$ this formula is 
equivalent to the residue
formula from \cite{KriW2}. We also mention the particular case 
of this formula as $z_i \to \infty$:
\beq\label{3d0}
\p^3_0 F=
\frac{1}{4\pi}\oint_{\gamma}
\frac{|dw(z)|^3}{\sigma (z ,\bar z)dz d\bar z}\,.
\eeq

\subsection{Conformal dynamics}
\label{sec:conformal}

Consider deformations such that the moments $t_k$ with
$k\geq 1$ are kept fixed and the only deformation parameter
is $t=t_0$ (``time'').
Equivalently, such deformations can be defined by the requirement that
the normal velocity of the boundary curve $\gamma$,
$V_n (z)=\lim_{\varepsilon \to 0}\delta_{\infty}n(z)/\varepsilon$,
at any point $z\in \gamma$ and at any time $t\in 
[t_1, \, t_2]$ with some $t_1<t_2$ is given by
\beq\label{cd1}
V_n(z)=\frac{\p_n \log |w(z)|}{2\sigma (z, \bar z)}=
\frac{|w'(z)|}{2\sigma (z, \bar z)}
\eeq
(see (\ref{deltainf})).
If $\sigma (z, \bar z)=\mbox{const}$, then equation
(\ref{cd1}) states that the normal
velocity of the interface $\gamma$ is proportional to the gradient
of the Green function of the Laplace operator. This is
the Darcy law for the dynamics of interface between viscous and
non-viscous fluids confined in the Hele-Shaw cell, assuming vanishing
surface tension at the interface:
$V_n(z)\propto |w'(z)|$. Such process is also called 
Laplacian growth (LG); see, e.g., 
\cite{RMP}-\cite{MWPT}.
For a non-constant $\sigma$ we have the
LG in a non-uniform background\footnote{For example,
the viscous flow
in the Hele-Shaw cell with a non-uniform spacing between the
glass plates.}.

The case when $\sigma (z, \bar z)$ is the squared modulus of a
holomorphic function is special and important. In this case
the problem can be mapped to another LG problem,
in some other plane which we call the $Z$-plane, with
uniform background but with
different boundary conditions. Namely, consider a map
$z\mapsto Z(z)$ which is conformal in the annulus ${\sf A}$.
The boundary curve $\gamma$ is mapped to a curve $\Gamma$ in the
$Z$-plane.
It is clear that the normal velocities, $V_n^{(z)}(z)$ and
$V_n^{(Z)}(Z)$, at the corresponding points of the
curves in the two planes are related as
$$
V_n^{(Z)}(Z)\Bigr |_{Z\in \Gamma}
=\left |\frac{dZ}{dz}\right | V_n^{(z)}(z)\Bigr |_{z\in \gamma}.
$$
Using (\ref{cd1}) and writing $\left | w'(z)
\right |=\left |\frac{dw}{dZ}\right |\cdot
\left |\frac{dZ}{dz}\right |$, we get
\beq\label{cd3}
V_n^{(Z)}(Z)=\frac{|dZ /dz |^2}{\sigma (z, \bar z)}\,
\bigl |\omega '(Z)\bigr |\,, \quad \quad \omega (Z):=w(z(Z)).
\eeq
Hence at $\sigma (z, \bar z) \propto |dZ /dz |^2$ we have
the LG problem in the $Z$-plane with the uniform background:
$V_n^{(Z)}(Z)\propto \bigl |\omega '(Z)\bigr |$.
Some examples are given in the next section.

{\bf Remark.}
In \cite{Z12}, for the particular case of the background density
$\sigma = R/|z|^2$ corresponding to the LG process on 
the surface of a cylinder $\{ Z \in \CC 
\, |\, 0\leq {\rm Im} Z\leq 2\pi R\}$
in the $Z$-plane,
the $z$-plane was called the {\it auxiliary physical plane} while
the $Z$-plane was called the {\it physical plane}.

\section{Symmetric solutions}

The case when the background
density function $\sigma$ (and the function $U$)
is {\it axially symmetric},
i.e., depends only on $|z|$, is of a special interest.
We study it in the rest of the paper.
In this case
we will write $U(z, \bar z)=U(z\bar z)$, $
\sigma (z, \bar z)=\sigma (z\bar z)$, etc and will sometimes denote
the argument of these functions of one variable by $x$
($x=|z|^2$). The corresponding solutions $F$ of equations
(\ref{eqs}) are called {\it symmetric} \cite{NZ13}. They are characterized
by the property that the derivatives $\p _k F$ restricted to zero 
values of the $t_k$'s with $k\geq 1$ vanish.

\subsection{Some general relations}

For any axially symmetric background
it holds:
\beq\label{t0}
z \p_z U(z\bar z)=z\bar z \, U'(z\bar z)=\bar z \p_{\bar z}U(z\bar z),
\eeq
\beq\label{t0a}
\sigma (x)=xU''(x)+U'(x)=\bigl (x U'(x)\bigr )',
\eeq
\beq\label{t0b}
\frac{1}{\pi}\int \!\! \int_{{\sf A}\cap {\sf D}} \sigma \, d^2z =
t_0 -r^2_0 U'(r_0^2).
\eeq
It is also easy to see that in expansions (\ref{series1}), (\ref{series2})
$\psi (z, \bar z)=0$ and
$$
u_0=r_0^2\log r_0^2\, U'(r_0^2)-U(r_0^2),
$$
so the expansions simplify and acquire the form
\beq\label{t0c}
\phi (z, \bar z)=\left \{
\begin{array}{l}
\displaystyle{-U(z\bar z)-u_0 +t_0\log (z\bar z) +
\sum_{k\geq 1}\bigl ( t_kz^k +\bar t_k \bar z^k\bigr ), \quad
z\in {\sf A}\cap {\sf D}}
\\ \\
\displaystyle{v_0 +\sum_{k\geq 1}\frac{1}{k}\bigl ( v_k z^{-k}+
\bar v_k \bar z^{-k}\bigr ), \quad \quad
z\in {\sf D^c}.}
\end{array}
\right.
\eeq

For symmetric solutions, formulas
(\ref{loc2}) and (\ref{comp2}) for the moments also simplify
because the contour integrals in their right hand sides vanish:
\beq\label{loc2a}
t_k=-\, \frac{1}{\pi k}\int \!\!\! \int_{{\sf D^c}\cap
{\sf B}(r_1)}\! z^{-k}
\sigma  (z\bar z) \, d^2 z\, ,
\quad \quad
v_k=\frac{1}{\pi}\int \!\!\int_{{\sf D}\cap {\sf B}(r_0)}\! z^{k}
\sigma  (z\bar z) \, d^2 z  , \quad \quad k\geq 1.
\eeq
Note
that these integrals do not depend on $r_1, r_0$
provided that $\gamma \subset {\sf A}={\sf B}(r_1)\setminus {\sf B}(r_0)$
and the function $\sigma$ is regular everywhere in
the annulus ${\sf A}$. In particular, if $\sigma$
is allowed to have singularities at $0$ and $\infty$ only, then
the integration can be extended to the whole domains
${\sf D^c}$ and ${\sf D}$ with the prescription that the
angular integration is performed first. Let us also mention
that the contour integral representation
(\ref{comp3a}) for $v_0$ in the symmetric case can be written
in the form
\beq\label{comp3b}
v_0=\frac{1}{2\pi i}\left (\oint_{\gamma} -
\oint_{|z|=r_0}\right ) \left (\log |z|^2 \p U - z^{-1} \, U\right )dz.
\eeq

\begin{proposition}\label{2F}
Suppose that only a finite number of the moments $t_k$ are different
from 0. Then the tau-function (\ref{F})
for a symmetric background can be represented as
\beq\label{Fv}
2F=-\, \frac{1}{\pi}\int \!\! \int_{{\sf A}\cap {\sf D}}
U\sigma \, d^2z +t_0v_0+
\sum_{k\geq 1}(t_k v_k+\bar t_k \bar v_k)-u_0t_0 +
u_0r_0^2 U'(r_0^2).
\eeq
\end{proposition}

\noindent
{\it Proof.}
Under the assumption of the proposition
expansion (\ref{t0c}) is
valid everywhere in ${\sf A}\cap {\sf D}={\sf D}\cap {\sf B}(r_0)$.
Substituting it into (\ref{F}) written in the form
$2F=\frac{1}{\pi}\int \!\!
\int_{{\sf A}\cap {\sf D}}\phi \, \sigma \, d^2z$,
performing the termwise integration and using the
definition of the complimentary moments, we get (\ref{Fv}).
\square

\noindent
{\bf Remark.} For any, not necessarily
symmetric, background the formula is basically the same;
only the $r_0$-dependent terms change (the last two terms in
(\ref{Fv})).

\begin{theorem}\label{lambda}
Let $U=U(z\bar z; \lambda)$ be a symmetric background
potential depending on a parameter $\lambda$.
Then the partial $\lambda$-derivative of $F$ taken at
constant $\{t_k\}^{\infty}_{0}$ is given by
\beq\label{Flambda}
\left. \frac{\p F}{\p \lambda}\right |_{\{t_k\}}=-\, \frac{1}{\pi}
\int \!\! \int_{{\sf A}\cap {\sf D}}
\p_{\lambda}U\, \sigma d^2z-
\, \left (t_0\! -\! r_0^2 U'(r_0^2)\right ) \p_{\lambda}u_{0},
\eeq
where $\p_{\lambda}u_{0}=
r_0^2 \log r_0^2 \p_{\lambda} U'(r_0^2) - \p_{\lambda} U (r_0^2)$.
\end{theorem}

\vspace{-2mm}

\noindent
{\it Proof.}
We should find the variation of $F=\frac{1}{2\pi}\int \!\!
\int_{{\sf A}\cap {\sf D}}\phi \, \sigma \, d^2z$ under the
deformation of the potential function
$U\to U +\delta U$, $\delta U=\p_{\lambda}U \delta \lambda $, and
a simultaneous deformation of the domain
that preserves the moments $t_k$ (see Proposition \ref{deltaU}).
We have:
$$
\delta F = \frac{1}{2\pi}\int \!\!
\int_{{\sf A}\cap {\sf D}}\! \delta \phi \, \sigma \, d^2z +
\frac{1}{2\pi}\int \!\!
\int_{{\sf A}\cap {\sf D}}\phi \, \delta \sigma \, d^2z +
\frac{1}{2\pi}\oint_{\gamma} \phi \, \delta n \, \sigma \, |dz|,
$$
where $\delta \phi = -\delta U -\delta u_0$
(this follows from (\ref{t0c}) at $\delta t_k=0$),
$\delta \sigma = \frac{1}{4}\, \Delta \delta U$ and $\delta n$ is
given by equation (\ref{delta1}). Plugging all this into the
right hand side and using the Green theorem in the form
$
\int \!\!
\int_{{\sf A}\cap {\sf D}}(f \Delta g - g\Delta f)d^2 z=
\left (\oint_{\gamma}-\oint_{|z|=r_0}\right )
(f\p_n g - g\p_n f)|dz|
$,
we arrive at the expression
$$
\begin{array}{ll}
\delta F=&\displaystyle{
-\frac{1}{2} \left (t_0\! -\! r_0^2 U'(r_0^2)\right )\delta u_0
-\frac{1}{\pi}\int \!\!
\int_{{\sf A}\cap {\sf D}} \!\! \delta U \, \sigma d^2z}
\\ & \\
&+\displaystyle{\,\, \frac{1}{8\pi}\oint_{|z|=r_0}\!\!
\left ( \delta U \, \p_n \phi \! -\! \phi \, \p_n \delta U \right )|dz|
+\, \frac{1}{8\pi}\oint_{\gamma}\!\!
\left (\phi \, \p_n  \delta U^H - \delta U  \p_n \phi \right )|dz|.
}
\end{array}
$$
The last integral is equal to zero because we can write
$\oint_{\gamma}\phi \, \p_n \delta U^H |dz|=
\oint_{\gamma} \delta U^H  \p_n  \phi |dz|$
(since both functions $\phi$ and
$\delta U^H$ are harmonic in ${\sf D^c}$) and then
the integrand vanishes on $\gamma$ since $\delta U^H=\delta U$ there.
One can also note that the terms like $t_kz^k$ and
$\bar t_k \bar z^k$ that are present in the expansion of $\phi$,
owing to the symmetry,
do not contribute to the integral over the circle $|z|=r_0$.
The remaining terms give the right hand side of equation (\ref{Flambda}).
\square

\vspace{-5mm}

\begin{corollary}\label{lambdaj}
Let $U(z\bar z)$ be a symmetric background
potential of the form
$$
U(z\bar z)=\sum_j \lambda_j U_j(z\bar z)
$$
with some functions $U_j$ and
parameters $\lambda_j$. Then
\beq\label{Flambda1}
\sum_j \lambda_j \frac{\p F}{\p \lambda_j}=
-\, \frac{1}{\pi}
\int \!\! \int_{{\sf A}\cap {\sf D}}\! U \sigma d^2z-
u_0t_0 +u_0 r_0^2 U'(r_0^2).
\eeq
\end{corollary}

\vspace{-3mm}

\noindent
This directly follows from equations (\ref{Flambda})
written for $\p_{\lambda_j}F$.

\begin{corollary} \label{cor:hom}
The function $F$ is a homogeneous function
of $t_0, \{t_k\}_{1}^{\infty}, \{\bar t_k\}_{1}^{\infty}$ and
$\lambda_j$'s of degree two:
\beq\label{Flambda2}
2F=\sum_j \lambda_j \p_{\lambda_j}F +t_0\p_0 F +
\sum_{k\geq 1}(t_k \p_k F +\bar t_k \bar \p_k F).
\eeq
\end{corollary}

\noindent
This is obtained by combining
(\ref{Flambda1}), (\ref{Fv}) and (\ref{vk}).

\subsection{The restriction 
to the $t_0$-line}

Let us call the line 
$t_k=0$ for all $k\geq 1$ (but $t_0\neq 0$)
{\it the $t_0$-line} in the space of moments\footnote{Depending
on the function $\sigma$, the values of $t_0$ corresponding to 
domains in the $z$-plane may belong to some interval of this line.}. 
For symmetric solutions, it is natural to consider the restriction 
of any quantity depending on the domain to the $t_0$-line.
Below we will denote such restriction by $(\ldots )\bigr |_{t_0}$.

The symmetry $U(z, \bar z)=U(z\bar z)$
implies that when all moments $t_k$ at $k\geq 1$ are
equal to 0,
the domain ${\sf D}$ is a disk of radius $r$ such that
\beq\label{t00}
t_0= \frac{1}{2\pi i}\oint_{|z|=r} \! z\p U \, \frac{dz}{z}=
r^2\, U'(r^2)
\eeq
(see (\ref{t0})).
Differentiating this equality w.r.t. $t_0$, we get:
\beq\label{t00a}
\p_{t_0}r^2(t_0) \, \sigma (r^2(t_0))=1 \quad \mbox{or}\quad
\p_{t_0}\log r^2 \bigr |_{t_0}=F'''(t_0)= 
\frac{1}{r^2 \sigma (r^2)}\,,
\eeq
where $F(t_0):=F\bigr |_{t_0}$ is 
the restriction of the function $F$ to the $t_0$-line.

The complimentary moments $v_k$ with $k\geq 1$ vanish,
$v_k \bigr |_{t_0}= \p_{k}F\bigr |_{t_0} =0$ for all
$k\geq 1$,
while $v_0$ is given by
\beq\label{v0sym}
v_0\bigr |_{t_0}=\int_{r_0^2}^{r^2}\!\! \log x \, \sigma (x)\, dx=
t_0 \log r^2 -U(r^2)-u_0.
\eeq
The potential $\phi$ is
\beq\label{phi0}
\phi (z, \bar z)\bigr |_{t_0}=\left \{
\begin{array}{l}
\displaystyle{-U(z\bar z)-u_0 +t_0\log (z\bar z), \quad \quad
r_0 \leq |z|\leq r}
\\ \\
\displaystyle{\,\,\, v_0,  \quad \quad \quad
|z|>r.}
\end{array}
\right.
\eeq
Since it depends on $|z|^2$ only, in this subsection
we will write
$\phi (z, \bar z)\bigr |_{t_0}=\phi (z\bar z)$.

There are several integral formulas for the function $F(t_0)$.
One of them is obtained by a direct calculation of the
integral (\ref{F}) in polar coordinates using the integral
$$
\int_{0}^{2\pi}\!  \log \left (
a^2 +b^2 \! -\! 2ab \cos \varphi \right )d\varphi =\pi \log \Bigl [
\mbox{max}\, \bigl ( a^2, b^2 \bigr )\Bigr ].
$$
This leads to the double integral formula
\beq\label{F0}
F(t_0)
=\int_{r_0^2}^{r^2}\!\! \sigma (x)dx 
\int_{r_0^2}^{x}\log x' \, \sigma (x')\, dx',
\eeq
which can be further simplified by taking into account that
$\log x \sigma (x)$ is a full derivative:
$\log x \sigma (x)=\p_x \Bigl ( U'(x)x\log x -U(x)\Bigr )$.
Then we obtain from (\ref{F0}):
\beq\label{F0b}
F(t_0)
=\int_{r_0^2}^{r^2}\!\Bigl ( U'(x)x\log x - U(x)\Bigr )\sigma (x) dx
-\left ( t_0 -r_0^2 U'(r_0^2)\right )u_0.
\eeq
Here $r^2$ should be understood as a function of $t_0$
implicitly given by (\ref{t0}). 
From this formula one can easily
see, using (\ref{t00a}), that $F'(t_0)=v_0$ and
$\p v_0 /\p t_0 =F''(t_0)=\log r^2$,
as it should be.
Another integral formula follows from (\ref{comp4}) and
(\ref{F}) and from (\ref{phi0}), with taking into account (\ref{t0b}):
\beq\label{F0a}
\begin{array}{lll}
2F(t_0) &=&\displaystyle{
\int_{r_0^2}^{r^2}\!\! \phi (x)
\sigma (x)dx}
\\ &&\\
&=&\displaystyle{\!\! \!\! -
\int_{r_0^2}^{r^2} U(x)\bigl (xU'(x)\bigr )'dx
+t_0(v_0- u_0)+u_0 r_0^2 U'(r_0^2)}
\\ &&\\
&=&\displaystyle{\!\! \int_{r_0^2}^{r^2} x\bigl (
U'(x)\bigr )^2 dx +t_0^2 \log r^2 \! - \! 2t_0 U(r^2) \! - \! 2u_0 t_0 +
r_0^4 \bigl (U'(r_0^2)\bigr )^2 \log r_0^2}.
\end{array}
\eeq
It is easy to check directly that (\ref{F0b}) and (\ref{F0a})
are equivalent.

{\bf Remark.}
We would like to mention that on the $t_0$-line the 
2D inverse potential problem for the Laplace operator $\Delta =4\p \bar \p$
becomes one-dimensional (for the operator $\p^2/\p X^2$).
To see this, we introduce the new 
variable $X=\log (z\bar z/r_0^2)\geq 0$
and denote ${\cal U}(X)=U(r_0^2 e^X)$.
Then equation (\ref{F0a}) can be rewritten as
\beq\label{F0c}
2F(t_0)=-\int_{\log r_0^2}^{\log r^2} {\cal U}(X)
{\cal U}''(X)dX +t_0 v_0 -\left (t_0 \! -\! 
{\cal U}'(\log r_0^2)\right )u_0.
\eeq
Set $\displaystyle{{\cal U}(X)=\sum_{k\geq 2} \tau_k X^k}$, then 
condition (\ref{t00}) (equivalent to the continuity of the 
derivative of the potential $\phi$ at $X=\log r^2$, i.e., 
$\p_X \left ( t_0 X-{\cal U}(X)\right ) \Bigr |_{X=\log r^2}=0$) reads
$$
t_0=\sum_{k\geq 2}k\tau_k (\log r^2)^{k-1}
$$
which means that $u=\log r^2 (t_0, \tau_2, \tau_3, \ldots )$ 
satisfies equations of the 
dispersionless KdV hierarchy
\beq\label{kdv}
\frac{\p u}{\p \tau_k}+k u^{k-1}
\frac{\p u}{\p t_0}=0, \quad k\geq 2.
\eeq

\subsection{Examples}

\subsubsection{The homogeneous density $\sigma (z\bar z)=
c(z\bar z)^{\alpha -1}$, $\alpha >0$}

Our first example is
\beq\label{examp1}
U(z\bar z)=\frac{c}{\alpha^2}\, (z\bar z)^{\alpha},
\quad \quad
\sigma (z\bar z)=
c(z\bar z)^{\alpha -1}
\eeq
with some real positive $c, \alpha$. We also have
\beq\label{examp1a}
z\p U (z\bar z) = \frac{c}{\alpha}\, (z\bar z)^{\alpha}=\alpha U (z\bar z).
\eeq

We start with the case of vanishing all the $t_k$'s except $t_0$:
\beq\label{examp2}
t_0=\frac{c}{\alpha}\, r^{2\alpha},
\quad \quad
v_0 \bigr |_{t_0}= \frac{t_0}{\alpha}\, 
\log\Bigl (\frac{\alpha t_0}{c}\Bigr )
- \frac{t_0}{\alpha} -u_0
\eeq
and
\beq\label{examp3}
F(t_0)= \frac{t_0^2}{2\alpha}\, \log\Bigl (\frac{\alpha t_0}{c}\Bigr )
-\frac{3t_0^2}{4\alpha}-u_0t_0 +c_0\,,
\eeq
where
$$
u_0= \frac{c}{\alpha}\, r_0^{2\alpha}\log r_0^2 -\frac{c}{\alpha^2}\, r_0^{2\alpha}, \quad 
c_0=\frac{c^2r_0^{4\alpha}}{4\alpha^3}
\bigr (2\alpha \log r_0^2 -1\bigr ).
$$
One can check that
$
F''(t_0)=\frac{1}{\alpha}\, \log\Bigl (\frac{\alpha t_0}{c}\Bigr )=
\log r^2
$.

Now let us consider the case of non-zero moments $t_k$.
At $\alpha >0$
all integrals converge at the origin and one can put $r_0=0$ but
we will keep it non-zero for illustrative purposes.

\begin{proposition} \label{alpha}
Assuming that only a finite number of the moments $t_k$
are different
from 0, the dispersionless tau-function for this solution
is quasi-homogeneous,
that is it obeys the relation
\beq\label{quasi-hom}
2F= t_0 \p_0 F +
\sum_{k\geq 1} \Bigl ( 1-\frac{k}{2\alpha}\Bigr )
\left ( t_k \p_k F +\bar t_k \bar \p_k F\right ) +Q(t_0),
\eeq
where
$\displaystyle{
Q(t_0)=-\frac{t_0^2}{2\alpha} -u_0 t_0 +2c_0}
$.
\end{proposition}

\vspace{-3mm}

\noindent
{\it Proof.}
We use equation (\ref{Fv}). In the integral term we write
$U\sigma = \bar \p (U\p U)-\p U \, \bar \p U$ and notice that for the
particular function $U$ we have $U\sigma =\p U \, \bar \p U$, and also
$z\p U=\alpha U$, so
$U\sigma = \frac{1}{2}\, \bar \p (U\p U)$. This allows us to
transform the 2D integral to a contour integral:
$$
\frac{1}{\pi}\int \!\! \int_{{\sf A}\cap {\sf D}} U \sigma \, d^2 z =
\frac{1}{4\pi i\alpha}\left (\oint_{\gamma}-
\oint_{|z|=r_0}\right )
(z\p U)^2 \, \frac{dz}{z}=
\frac{1}{4\pi i\alpha}\oint_{\gamma}
(z\p U)^2 \, \frac{dz}{z} -\frac{c^2r_0^{4\alpha}}{2\alpha^3}\,.
$$
Now recall (\ref{jump}) and represent $\p U=C^+-C^-$ (on $\gamma$),
with $C^+$ being a polynomial.
Shrinking the integration contour to $\infty$, we obtain:
\beq\label{same}
\frac{1}{2\pi i}\oint_{\gamma} (z\p U)^2 \, \frac{dz}{z} =
t_0^2 +2\sum_{k\geq 1}kt_k v_k.
\eeq
Since the initial integral is obviously real, we conclude that
$\displaystyle{\sum_{k\geq 1}kt_k v_k}$ is a real quantity, i.e.,
$ \displaystyle{\sum_{k\geq 1}kt_k v_k=\sum_{k\geq 1}k\bar t_k \bar v_k}$.
The quasi-homogeneity
relation (\ref{quasi-hom}) follows.
\square

\noindent
Note that the function $\tilde F = F-F(t_0)$ satisfies the same
relation (\ref{quasi-hom}) without the last term $Q(t_0)$.

The following formulas are direct consequences of Corollary \ref{cor:hom}
and equation (\ref{quasi-hom}):
\beq\label{examp4}
2F=c\p_c F+ t_0\p_0 F + \sum_{k\geq 1}(t_k \p_k F+\bar t_k \bar \p_k F),
\eeq
\beq\label{examp5}
c\p_c F= -\frac{1}{\alpha}\sum_{k\geq 1}kt_k \p_k F +Q(t_0).
\eeq

\begin{figure}[tb]
\epsfysize=8.0cm
\centerline{\epsfbox{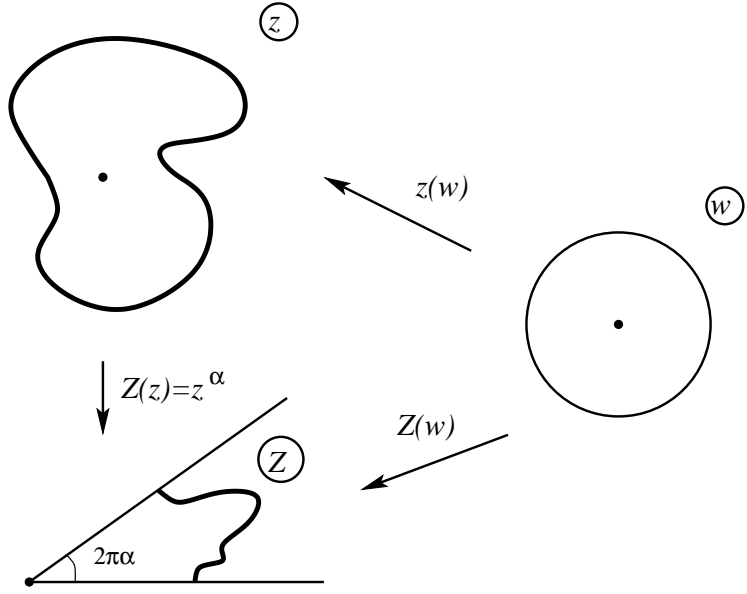}}
\caption{\sl The conformal dynamics with $\sigma 
=c (z\bar z)^{\alpha -1}$
is mapped to the Laplacian growth in the cone with the angle $2\pi \alpha$.}
\label{fi:cone}
\end{figure}

The case $\alpha =1$ ($\sigma (z, \bar z)=\mbox{const}$)
was considered in \cite{MWZ,WZ} in connection with the
Laplacian growth
(the Hele-Shaw problem) in the plane with a sink at infinity.
As is mentioned in \cite{MWZ99}, at arbitrary
$0< \alpha \leq 1$ the conformal dynamics
can be mapped
to a LG process in a sector with angle
$2\pi \alpha$ and periodic conditions at the boundary rays (a cone),
see Fig.~\ref{fi:cone}. Some details are given below.

We map the whole $z$-plane punctured at $z=0$ to the sector
$${\sf S}_{\alpha}=
\Bigl \{ Z\in \CC \setminus \{0\} \Bigr |\,
0\leq \mbox{arg}\, Z<2\pi \alpha \Bigr \}$$
in the $Z$-plane by the map
$Z(z)=z^{\alpha}$. Let ${\sf D}^{\pm }_{\alpha}$
be the images of ${\sf D^c}$
and ${\sf D}$ respectively. The image of the curve $\gamma$ is
a curve $\Gamma \subset {\sf S}_{\alpha}$ such that its endpoints
on the boundary rays of the cone are at the same distance from the
origin. The moments $t_k$ and $v_k$ can be represented as integrals
in the $Z$-plane:
$$
t_0 = \frac{c}{2\pi i \alpha^2}\oint_{\Gamma}\bar Z  dZ=
\frac{c}{\pi  \alpha^2}\, \mbox{Area}\, \bigl ({\sf D}^-_{\alpha} \bigr ),
\quad \quad
t_k=-\frac{1}{\pi k}\int \!\! \int_{{\sf D}^+_{\alpha}}\! Z^{-k/\alpha}d^2Z,
$$
$$
v_0=\frac{c}{\pi \alpha^3}\int \!\! \int_{{\sf D}^-_{\alpha}}\!
\log |Z|^2 d^2Z, \quad \quad
v_k=\frac{c}{\pi \alpha^2}\int \!\! \int_{{\sf D}^-_{\alpha}}\!
Z^{k/\alpha}d^2Z.
$$
At $\alpha >0$ we put $r_0=0$ and
$r_1 \to \infty$ because all 2D integrals are either
convergent or can be made such by using the simple
additional prescription that the angular integration is made first.

The LG problem in ${\sf S}_{\alpha}$ is the following
moving boundary value problem:
\beq\label{D2}
\left \{
\begin{array}{l}
\Delta \Phi (Z, \bar Z)=0 \quad \mbox{in} \,\,\,\, {\sf D_+}
\\ \\
\Phi (Ze^{2\pi i \alpha}, \bar Z e^{-2\pi i \alpha})=\Phi (Z, \bar Z)
\\ \\
\Phi (Z, \bar Z)=0, \quad Z\in \Gamma
\\ \\
\Phi (Z, \bar Z)=-\, \frac{1}{2}\, \log |Z| +\ldots \quad \mbox{as} \quad
|Z| \rightarrow +\infty \,,
\end{array} \right.
\eeq
with the normal velocity of the curve $\Gamma$ given by
$V_n (Z)=-\p_n \Phi (Z, \bar Z)$, $Z\in \Gamma$. The conformal map
from the exterior of the unit circle in the $w$-plane onto
${\sf D}^+_{\alpha}$ is
$$\omega (Z)=w(Z^{1/\alpha})$$ (see section \ref{sec:conformal}).
Then $\Phi (Z, \bar Z)=-\frac{\alpha}{2}\, \log |\omega (Z)|$
meets all the requirements (\ref{D2}) and hence is the (unique) solution.
The normal velocity is then
$$
V_n (Z)=\frac{\alpha}{2}\,\p_n \log |\omega (Z)| =
\frac{\alpha}{2}\, |\omega '(Z)|.
$$
The moment $t_0$ is proportional to time while the moments
$t_k$ at $k\geq 1$ are Richardson's conserved quantities for this
problem \cite{Richardson}.

\subsubsection{The homogeneous density $\sigma (z\bar z)=
c(z\bar z)^{\alpha -1}$, $\alpha <0$}

With small modifications, the formulas given above hold in
the case $\alpha <0$ (and, in particular,
$\alpha =-1$).
The conformal dynamics in the $z$-plane 
can be mapped by the map $Z(z)=z^{\alpha}$
to the Laplacian growth in the compact
{\it interior} domain bounded by the curve $\Gamma$ and the 
rays $\mbox{arg} Z=0$,  $\mbox{arg} Z=2\pi \alpha$ which is 
now the image of the exterior domain ${\sf D^c}$ in the 
$z$-plane.

At negative $\alpha$ one can not put $r_0=0$.
Formulas for $t_k$ and $v_k$ remain the same at $k\geq 1$ 
but change at $k=0$:
$$
t_0=-\, \frac{c}{\pi  \alpha^2}\, 
\mbox{Area}\, \bigl ({\sf D}^+_{\alpha} \bigr ),
\quad \quad
v_0=\frac{c}{\pi \alpha^3}\int \!\! \int_{\tilde {\sf D}^-_{\alpha}}\!
\log |Z|^2 d^2Z.
$$
The domain $\tilde {\sf D}^-_{\alpha} \subset {\sf S}_{\alpha}$
is bounded by the curve $\Gamma$ and the arc $2\pi \alpha \leq
\mbox{arg}\, Z <0$.
This ``cut-off'' is necessary
because the domain ${\sf D}^-_{\alpha}$ is non-compact and the 
integral over the whole ${\sf D}^-_{\alpha}$ diverges.
Note that $t_0$ becomes negative and the physical time 
is $t\propto -t_0$.

\subsubsection{The homogeneous density $\sigma (z\bar z)=
R/(z\bar z)$}

Formally, this is the limiting case of the previous family
of solutions with $\alpha \to 0$. However, because the limit
is not easy to perform and, most important, because this case
is very interesting by itself due to the connection with
Hurwitz numbers, it deserves a separate consideration.
It is convenient
to choose the function $U(z\bar z)$ in such a way that
$U(r_0)=U'(r_0)=0$:
\beq\label{h2}
U=\frac{R}{2} \Bigl [ \log \frac{z\bar z}{r_0^2}\Bigr ]^2,
\quad
z\p U=R\log \frac{z\bar z}{r_0^2}\,,
\quad
\p \bar \p U =\sigma =\frac{R}{z\bar z}
\eeq
where $R$ is a parameter of the solution. Sometimes
we will also use the parameter $\beta =1/R$.

Again, we start with the case of the vanishing $t_k$'s
except $t_0$ (the $t_0$-line):
\beq\label{hur1}
t_0=R\log \frac{r^2}{r_0^2}, \quad \quad
v_0 \bigr |_{t_0}
=t_0\log r^2 -\frac{R}{2} \Bigl [ \log \frac{r^2}{r_0^2}\Bigr ]^2
=\frac{t_0^2}{2R}+t_0 \log r_0^2
\eeq
and $u_0=0$, so
\beq\label{hur2}
F(t_0)=\frac{t_0^3}{6R}+\frac{t_0^2}{2}\, \log r_0^2.
\eeq
One can check that $F''(t_0)=\beta t_0 +\log r_0^2 =\log r^2$.

Now we address the general case of non-zero $t_k$'s.

\begin{proposition} \label{alpha=0}
The dispersionless tau-function for the solution
determined by the data (\ref{h2}) obeys
the following homogeneity relation:
\beq\label{hom}
2F= R \p_{R}F +t_0\p_0 F +
\sum_{k\geq 1}
\left ( t_k \p_k F +\bar t_k \bar \p_k F\right )
\eeq
(the $R$-derivative is taken at constant $t_k$'s).
\end{proposition}

\noindent
This is a particular case of (\ref{Flambda2}) (see Corollary
\ref{cor:hom}).

\begin{proposition} \label{cut-and-join}
The dispersionless tau-function (\ref{hom}) satisfies the relations
\beq\label{h3}
\begin{array}{l}
\displaystyle{\frac{\p F}{\p \log r_0^2}=
\frac{t_0^2}{2}+\sum_{k\geq 1}kt_k \, \p_kF},
\\ \\
\displaystyle{\frac{\p F}{\p \beta}=\frac{t_0^3}{6}+
t_0\sum_{k\geq 1}kt_k \, \p_kF +\frac{1}{2}
\sum_{k,l\geq 1} \bigl ( kl t_kt_l \, \p_{k+l}F+
(k+l)t_{k+l}\, \p_k F \, \p_l F\bigr )},
\end{array}
\eeq
where the derivatives are taken at constant $t_0, t_1, t_2 , \ldots$
\end{proposition}

\noindent
{\it Proof.} The first formula can be proved by a direct variation
of the function $F$ under a small change
$r_0^2 \to r_0^2 +\delta r_0^2$
similar to the one done in the proof of
Theorem \ref{lambda}.
On the one hand, the calculation in the present case
is somewhat
simpler because the corresponding $\delta U$ is harmonic in
${\sf A}\cap {\sf D}$ but, on the other hand, one should take into account
that the interior boundary of the domain (the circle $|z|=r_0$)
also moves. An accurate calculation which we omit here
gives:
$$
\frac{\p F}{\p \log r_0^2}=Rv_0 -Rt_0 \log r_0^2.
$$
(see also \cite{Z12} for a different proof).
To proceed, we use (\ref{comp3b}) and the fact that our particular
potential $U$ satisfies the relation $U=(z\p U)^2 / (2R)$,
so we can write
$$
v_0=\frac{1}{2\pi i}\left (\oint_{\gamma}-\oint_{|z|=r_0}\right )
\left (\log |z|^2 z\p U - U\right ) \frac{dz}{z}=
t_0 \log r_0^2 +\frac{1}{4\pi i R}\oint_{\gamma}(z\p U)^2 \frac{dz}{z}\,.
$$
Then, using the same argument as in the proof of
Proposition \ref{alpha} (see equation (\ref{same})),
we get the first equation in (\ref{h3}).

For the proof of the second formula we note that
in the case (\ref{h2}) $U \sigma = \frac{1}{3}\, \bar \p (U\p U)$
and thus
$$
\frac{1}{\pi}\int \!\! \int_{{\sf A}\cap {\sf D}}U\sigma \, d^2z =
\frac{\beta}{12\pi i}\oint_{\gamma}(z\p U)^3 \, \frac{dz}{z}
$$
(the integral over the interior boundary is absent because
$\p U=0$ there).
Proceeding as in the proof of Proposition \ref{alpha}, we obtain
$$
\frac{1}{12\pi i}\oint_{\gamma}(z\p U)^3 \, \frac{dz}{z}=
\frac{t_0^3}{6}+
t_0 \! \sum_{k\geq 1}kt_k \, v_k +\frac{1}{2}
\sum_{k,l\geq 1} \bigl ( kl t_kt_l  v_{k+l}+
(k+l)t_{k+l} v_k  v_l\bigr ).
$$
Then from Proposition \ref{2F} we have:
\beq\label{vn6b}
\begin{array}{ll}
2F &= \displaystyle{
t_0v_0 \! +\! \sum_{k\geq 1}(t_kv_k +\bar t_k \bar v_k)
-\frac{\beta t_0^3}{6}-\beta t_0 \! \sum_{k\geq 1} kt_kv_k}
\\ &\\
&\, -\,\,\, \displaystyle{\frac{\beta}{2}\sum_{k,l\geq 1}
\Bigl (kl t_k t_l v_{k+l}+ (k+l)t_{k\! +\! l}v_kv_l\Bigr ).}
\end{array}
\eeq
Combining this with the homogeneity
property (Proposition \ref{alpha=0}), we arrive at
the second formula in (\ref{h3}).
\square

\begin{figure}[tb]
\epsfysize=8.0cm
\centerline{\epsfbox{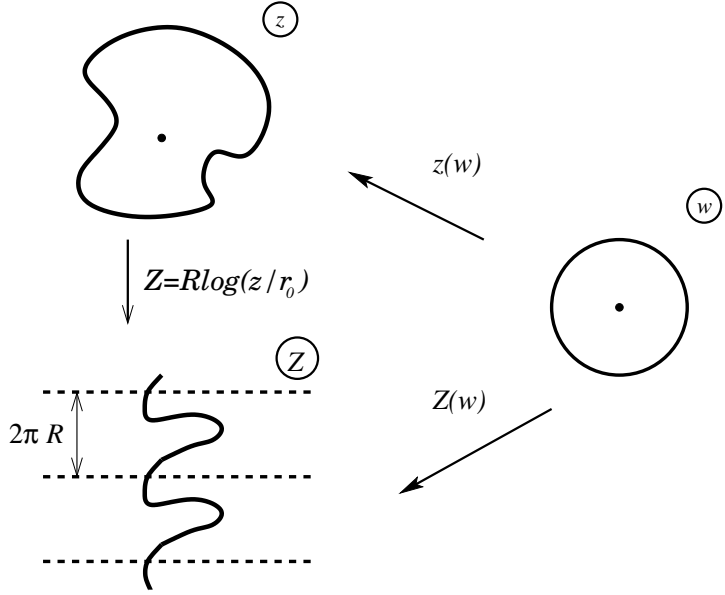}}
\caption{\sl The conformal dynamics with $\sigma =R/|z|^2$
is mapped to the Laplacian growth on the cylinder of 
radius $R$. (In the $Z$-plane, two copies of the curve 
$\Gamma$ are shown.)}
\label{fi:cyl}
\end{figure}

\vspace{-3mm}

\noindent
{\bf Remark.}
Combinatorial formulas
for Taylor expansion coefficient of the function $F$ have been suggested in
\cite{NZ13}. These coefficients are essentially the double Hurwitz
numbers for connected genus 0 coverings of the Riemann sphere.
The double sum in the second equation  in (\ref{h3}) is the
genus 0 part of the celebrated cut-and-join operator \cite{GJ}, see also \cite{Takasaki12}.
At the same time, this function is closely connected with the
Laplacian growth problem on a cylinder, see below and \cite{Z12}.

As is shown in \cite{Z12}, the conformal dynamics in the $z$-plane
punctured at $z=0$
with $\sigma = R/|z|^2$ can be mapped to a LG process on the surface 
of an infinite cylinder of radius $R$, see Fig. \ref{fi:cyl}.
We map the whole $z$-plane to the strip $${\sf C}_R=
\{ Z\in \CC \setminus \{0, \infty\} \, \bigr | 0\leq {\rm Im}\, Z <2\pi R\}$$
in the $Z$-plane by the map $Z(z)=R\log (z/r_0)$. Let 
${\sf D}^{\pm}$ be the images of ${\sf D^c}$ and ${\sf D}$ 
respectively. The image of the curve $\gamma$ is a curve 
$\Gamma \subset {\sf C}_R$ such that the endpoints on the 
two sides of the strip have the same real parts. The moments
$t_k$ and $v_k$ can be represented as integrals in the $Z$-plane:
$$
t_0 =
\frac{1}{2\pi i R}\int_{\Gamma}(Z \! +\! \bar Z)dZ =
\frac{1}{\pi R}\int_{\Gamma}XdY =
\frac{\mbox{Area}({\sf D}_{-}^{(0)} )}{\pi R},
$$
$$
v_0=\frac{R}{\pi}\int \!\!\!\int_{{\sf D}\setminus 
{\sf B}(r_0)}\!\frac{\log (z\bar z)}{z\bar z}\, d^2 z \, =\,
2t_0\log r_0 +\frac{2}{\pi R^2}\int \!\!\! \int_{{\sf D}_{-}^{(0)}}
Xd^2Z\,,
$$
$$
t_k=
-\, \frac{r_0^{-k}}{\pi k R}\int \!\!\! \int_{{\sf D_+}}
\! e^{-kZ/R}d^2Z, \quad \quad
v_k=
\frac{r_0^{k}}{\pi  R}\int \!\!\! \int_{{\sf D_-}}
\! e^{kZ/R}d^2Z.
$$
Here ${\sf D}_{-}^{(0)}$ is the 
domain bounded by
the curve $\Gamma$ and the section ${\rm Re}\, Z =0$. 
Note that ${\sf D}_{-}^{(0)}$ is the image of 
${\sf B}(r_0)$ under the map from the auxiliary physical plane.

The LG problem in ${\sf C}_R$ is the following moving
boundary value problem:
\beq\label{D2aaa}
\left \{
\begin{array}{l}
\Delta \Phi (Z, \bar Z)=0 \quad \mbox{in} \,\,\,\, {\sf D_+}
\\ \\
\Phi (Z+2\pi iR, \bar Z -2\pi iR)=\Phi (Z, \bar Z)
\\ \\
\Phi (Z, \bar Z)=0, \quad Z\in \Gamma
\\ \\
\Phi (Z, \bar Z)=-\, \frac{1}{2}\, {\rm Re}\, Z +
\ldots \quad \mbox{as} \quad
{\rm Re}\, Z \rightarrow +\infty \,,
\end{array} \right.
\eeq
with the normal velocity of the curve $\Gamma$ given by
$V_n (Z)=-\p_n \Phi (Z, \bar Z)$, $Z\in \Gamma$.
The conformal map
from the exterior of the unit circle in the $w$-plane onto
${\sf D}^+$ is
$\omega (Z)=w\left (r_0e^{Z/R}\right )$.
Then $\Phi (Z, \bar Z)=-\frac{R}{2}\, \log |\omega (Z)|$
meets all the requirements (\ref{D2aaa}) and hence is the (unique) solution.
The normal velocity is then
$$
V_n (Z)=\frac{R}{2}\,\p_n \log |\omega (Z)| =
\frac{R}{2}\, |\omega '(Z)|.
$$
The moment $t_0$ is proportional to time while the moments
$t_k$ at $k\geq 1$ are Richardson's conserved quantities for this
problem \cite{Richardson}.

\subsubsection{A more general family of examples}

All previous examples can be unified in a broader
family by setting
\beq\label{other1}
U(z\bar z)=\left ( C_1 \log (z\bar z) + C_0 \right )^{\nu}
\eeq
with some constants $C_1$, $C_0$ and $\nu =\frac{k-1}{k-2}$
with integer $k> 2$, so
\beq\label{other2}
z \p U= C_1 \nu \left ( C_1 \log (z\bar z) + C_0 \right )^{\nu -1},
\quad
\sigma = \frac{C_1^2 \nu (\nu \! -\! 1)}{z\bar z}
\left (C_1 \log (z\bar z) + C_0 \right )^{\nu -2}.
\eeq
All functions $U$ of the form (\ref{other1})
satisfy the differential equation
\beq\label{other3}
U''(x)-\frac{1}{k\! -\! 1}\, \frac{(U'(x))^2}{U(x)}+\frac{1}{x}\, U'(x)=0,
\eeq
with the first integral
$C_1 \nu \, U^{\frac{1}{k-1}}=U'$ which can be also written as
\beq\label{other4}
U=\left (C_1\nu \right )^{1-k} \Bigl (z\p U\Bigr )^{k-1}.
\eeq
We assume that $r_0^2 >e^{-C_0/C_1}$, so these functions
are non-singular in ${\sf A}$.

The case $k=3$ at $C_1=\sqrt{R/2}$, $C_0 =-\sqrt{R/2}\, \log r_0^2$
is the potential (\ref{h2}).
Formally, the case $k=2$ also belongs to this family
and corresponds to potential (\ref{examp1})
since
$$
\lim_{k\to 2} \Bigl (\alpha (k\! -\! 2)
\log (z\bar z) +1\Bigr )^{\frac{k-1}{k-2}}=(z\bar z)^{\alpha}.
$$
At $k=4$ the potential is
$
U(z\bar z)=\bigl ( C_1 \log (z\bar z) + C_0 \bigr )^{3/2}.
$

At the $t_0$-line we have:
$$
\begin{array}{l}
\displaystyle{t_0=C_1\nu \left ( C_1 \log r^2 +C_0\right )^{\nu -1}},
\\ \\
\displaystyle{v_0 \bigr |_{t_0}=(\nu -1)
\left (\frac{t_0}{C_1 \nu}\right )^{\frac{\nu}{\nu -1}} -
\frac{C_0}{C_1}\, t_0 -u_0,}
\\ \\
\displaystyle{u_0=\left (C_1(\nu \! -\! 1)\log r_0^2 -C_0\right )
\left ( C_1 \log r_0^2 +C_0\right )^{\nu -1}.
}
\end{array}
$$
The calculation of $F(t_0)$ by means of (\ref{F0a}) gives:
$$
F(t_0)= C_1 \frac{\nu (\nu -1)^2}{2\nu -1}
\left (\frac{t_0}{C_1\nu}\right )^{\frac{2\nu -1}{\nu -1}}
-\frac{C_0}{2C_1} t_0^2 -u_0 t_0 +K_0,
$$
where
$$
K_0=\frac{C_1^2 \nu ^2}{2\nu -1}
\left ( C_1 \log r_0^2 +C_0\right )^{2\nu -2}\left [
(\nu -1)\log r_0^2 -\frac{C_0}{2C_1}\right ]
$$
or
\beq\label{other5}
F(t_0)=\frac{a_k}{C_1^{k-1}}\, t_0^k
-\frac{C_0}{2C_1} t_0^2 -u_0 t_0 +K_0\,,
\quad \quad a_k= \frac{1}{k}\,
\frac{(k-2)^{k-2}}{(k-1)^{k-1}}.
\eeq
It is easy to check that $\displaystyle{F''(t_0)=k(k\! -\! 1)a_k
\frac{t_0^{k-2}}{C_1^{k-1}}-
\frac{C_0}{C_1}= \log r^2}$ as it should be.

In order to make this family of examples closer to the formulas
for the case $k=3$ we set
$
R_k :=(k-1)C_1^{k-1}$, 
$C_0=-C_1 \log r_0^2\,
$
and
$$
U_k(z\bar z)=\left (\frac{R_k}{k-1}\right )^{\frac{1}{k-2}}
\left ( \log \frac{z\bar z}{r_0^2}\right )^{\frac{k-1}{k-2}},
$$
so that $U_3$ coincides with 
$U$ from equation (\ref{h2}) with $R=R_3$. We also have
$U_k(r_0^2)=U_k'(r_0^2)=0$.
Equation (\ref{other5})
acquires the form
\beq\label{other6}
F(t_0)=\frac{(k\! -\! 1)a_k}{R_k}\, t_0^k +\frac{t_0^2}{2}\, \log r_0^2.
\eeq
In the general case of $t_k\neq 0$
we can write, taking into account (\ref{other4}):
$$
\frac{1}{\pi}\int \!\! \int_{{\sf A}\cap {\sf D}}\! U_k \sigma _kd^2z =
\frac{(k\! -\! 1)(k\! -\! 2)a_k}{2\pi i \, R_k}\oint_{\gamma}
(z\p U_k)^k \frac{dz}{z}\,.
$$
This relation allows one to obtain analogs of equation (\ref{vn6b})
with higher $W_k$-operators instead of the cut-and-join operator
$W_3$. 

One can also consider the potential $\sum_k U_k(z\bar z)$.
The corresponding function $F$ is presumably the dispersionless limit
of the tau-function
(or rather the free energy) for the Gromov-Witten theory 
of $\CC P^1$ \cite{OP02}.

\section*{Acknowledgments}

\addcontentsline{toc}{section}{Acknowledgments}

Discussions with A.Alexandrov, I.Krichever, A.Morozov and S.Na\-tan\-zon
are gratefully acknowledged.
Some of these results were reported 
at the 2nd International Workshop on Nonlinear and 
Modern Mathematical Physics (March 9-11, 2013, USF, Tampa, Florida).
The author thanks the organizers and especially professors
W.-X.Ma and R.Teodorescu for the invitation and support.
This work was supported in part
by RFBR grant 12-01-00525, by joint RFBR grants 12-02-91052-CNRS,
12-02-92108-JSPS, by grant NSh-3349.2012.2 for support of
leading scientific schools and
by Ministry of Science and Education of Russian Federation
under contract 8207.

\end{document}